\documentclass[aps,twocolumn,groupedaddress,showpacs,showkeys,nofootinbib]{revtex4}

\def\slashchar#1{\setbox0=\hbox{$#1$}
   \dimen0=\wd0 \setbox1=\hbox{/} \dimen1=\wd1
   \ifdim\dimen0>\dimen1 \rlap{\hbox to \dimen0{\hfil/\hfil}} #1
   \else  \rlap{\hbox to \dimen1{\hfil$#1$\hfil}} / \fi}

\newcommand{\A}{{\overline A}}
\newcommand{\F}{{\overline F}}
\newcommand{\V}{{\overline V}}
\newcommand{\D}{{\bf D}}
\newcommand{\oS}{{\overline S}}
\newcommand{\oP}{{\overline P}}

\def\({{\Bigl(}}
\def\){\Bigr)}
\def\]{\Bigr]}
\def\[{\Bigl[}

\newcommand{\Nabla}{\overline{\nabla}}

\newcommand{\ga}{{g_A}}

\newcommand{\Det}{{\textrm {Det}}}
\newcommand{\tr}{{\textrm {tr}}}

\newcommand{\I}{{\cal I}}
\newcommand{\cL}{{\cal L}}

\begin{document}

\title{The Energy Momentum Tensor of Chiral Quark Models at Low Energies}

\author{E. Meg\'{\i}as}
\email{emegias@ugr.es}

\author{E. \surname{Ruiz Arriola}}
\email{earriola@ugr.es}

\author{L. L. Salcedo}
\email{salcedo@ugr.es}

\affiliation{
Departamento de F\'{\i}sica Moderna,
Universidad de Granada,
E-18071 Granada, Spain}

\date{\today} 

\begin{abstract}
The low energy structure of the energy momentum tensor in several
chiral quark models with SU(3) flavor symmetry is analyzed in the
presence of external gravitational fields. To this end a derivative
expansion in a curved four dimensional space time is considered. This
allows us to compute the chiral coefficients $L_{11}$, $L_{12}$, and
$L_{13}$, appearing when the model is coupled to gravity and related
to space-time curvature properties.
\end{abstract}



\maketitle

\section{INTRODUCTION}
\label{sec:intro}

The energy momentum tensor plays a crucial role in quantum field
theory since it arises as a Noether current of the Poincare group. It
is conserved in all relativistic local theories even when there are no
other conserved charges.  According to the standard Noether
construction (see e.g.  Refs.~\cite{GamboaSaravi:2003aq,
Forger:2003ut} for recent reviews on the classical theory), the energy
momentum tensor can be computed as a functional derivative of the
action with respect to an external space-time-dependent metric, $
g_{\mu\nu} (x)$, around the flat space-time metric $ \eta_{\mu \nu} $
(we take the signature $(+ - - -)$ ).
\begin{eqnarray}
\frac{1}{2} \theta^{\mu \nu} (x) &=& \frac{\delta S}{\delta g_{\mu
\nu}(x)} 
\Big|_{g_{\mu \nu} = \eta_{\mu\nu} } 
\label{eq:theta_quark}
\end{eqnarray} 
where 
\begin{eqnarray}
S = \int d^4 x \sqrt{-g} \;{\cal L}(x)\,.   
\end{eqnarray} 
The chiral Lagrangian ${\cal L}$ contains only metric contributions,
and it is obtained by promoting the flat space-time metric $ \eta_{\mu
\nu} $ to the curved one $ g_{\mu \nu} $. It takes the form given in
Refs. \cite{Gasser:1983yg,Gasser:1984gg}. In QCD, the energy momentum
tensor operator probes the interaction of quarks and gluons to
gravitons. At the quantum level the high energy behavior of
$\theta_{\mu \nu} $ is improved if suitable additional transverse
corrections are implemented~\cite{Callan:1970ze}, and within such a
set up there is a trace anomaly relating $ \theta_\mu^\mu $ to the
divergence of the dilatation
current~\cite{Chanowitz:1972da,Collins:1976yq,Fujikawa:1980rc},
signaling the anomalous breaking of scale invariance. The
non-vanishing vacuum expectation value, $ \langle 0 | \theta_\mu^\mu |
0 \rangle $ is related to the existence of a non-vanishing gluon
condensate generating a bag constant~\cite{Ioffe:2002be} and scale
Ward identities \cite{Novikov:1981xj,Shifman:1988zk}. Deep inelastic
scattering provides also some information on the momentum fraction
carried by quarks and gluons in a hadron at a given
scale~\cite{Ji:1995sv}. Direct experimental determination on the basis
of one graviton exchange is out of question due to the smallness of
the gravitational constant as compared to weak and strong processes.
The gravitational pion form factor can be used to determine the width
of a light Higgs boson into two pions~\cite{Donoghue:1990xh}, and also
the decay of a scalar glueball into two
pions~\cite{Voloshin:1980zf}. There have also been few attempts in the
past to define $\theta_{\mu\nu} $ on the
lattice~\cite{Caracciolo:1989pt}, but so far there are no results of
practical interest concerning matrix elements between hadron states
carrying different momentum.

At low energies, the spontaneous breaking of chiral symmetry dominates
and in the meson sector any operator can be described in terms of a
non-linearly transforming Pseudoscalar Goldstone boson field $U$ with
an infinite number of low energy constants
(LEC's)~\cite{Langacker:1973hh,Weinberg:1978kz,
Gasser:1983yg,Gasser:1984gg,Donoghue:1991qv} (for review see, {\em e.g.},
Refs.~\cite{Donogue:1992bk,Pich:1995bw}). In a chiral expansion, the most general
structure of $\theta_{\mu\nu}$ up to and including fourth order
corrections reads~\cite{Donoghue:1991qv}
\begin{eqnarray}
\theta_{\mu \nu} = \theta_{\mu \nu}^{(0)}+\theta_{\mu \nu}^{(2)} +
\theta_{\mu \nu}^{(4)} + \dots 
\end{eqnarray}
with
\begin{eqnarray}
\theta_{\mu \nu}^{(0)} &=& -\eta_{\mu \nu} {\cal L}^{(0)} , \\
\theta_{\mu \nu}^{(2)} &=& \frac{f^2}4 \langle D_\mu U^\dagger D_\nu U
\rangle - \eta_{\mu\nu} {\cal L}^{(2)} ,
\label{eq:en-mom} \\ 
\theta_{\mu \nu}^{(4)} &=& -  \eta_{\mu\nu}{\cal
L}^{(4)} + 2 L_4 \langle D_\mu U^\dagger D_\nu U \rangle \langle
\chi^\dagger U + U^\dagger \chi \rangle \nonumber \\ &+& L_5 \langle
D_\mu U^\dagger D_\nu U + D_\nu U^\dagger D_\mu U \rangle \langle
\chi^\dagger U + U^\dagger \chi \rangle \nonumber \\ &-& 2
L_{11}\left( \eta_{\mu \nu} \partial^2 - \partial_\mu \partial_\nu
\right) \langle D_\alpha U^\dagger D^\alpha U \rangle \nonumber \\ &-&
2 L_{13} \left(\eta_{\mu \nu} \partial^2 - \partial_\mu \partial_\nu
\right) \langle \chi^\dagger U + U^\dagger \chi \rangle \nonumber \\
&-& L_{12} \left( \eta_{\mu\alpha} \eta_{\nu \beta} \partial^2 +
\eta_{\mu\nu} \partial_\alpha \partial_\beta - \eta_{\mu \alpha}
\partial_\nu \partial_\beta - \eta_{\nu \alpha} \partial_\mu
\partial_\beta \right) \nonumber \\ && \times \langle D^\alpha
U^\dagger D^\beta U \rangle \,, \nonumber \\
\end{eqnarray} 
where  $\langle A \rangle = \tr A $ means trace in flavor space
and the expansion for the Lagrangian reads  
\begin{eqnarray} 
{\cal L} &=& {\cal L}^{(0)}+{\cal L}^{(2)} +{\cal L}^{(4)} + \dots
\label{eq:chl_flat}
\end{eqnarray} 
The chiral expansion corresponds to an expansion in powers of external momenta. The upper-script means the order of the contribution. The zeroth order vacuum contribution reads
\begin{eqnarray} 
{\cal L}^{(0)} = B  \, , 
\label{eq:chi0_flat} 
\end{eqnarray} 
where $B$ is the vacuum energy density constant. The second
order contributions reads
\begin{eqnarray} 
{\cal L}^{(2)}  &=& {f^2\over 4} \langle D_\mu U^\dagger D^\mu U
+(\chi^\dagger U + U^\dagger \chi) \rangle ,
\label{eq:chl2_flat}  
\end{eqnarray} 
and the fourth order contribution is
\begin{eqnarray} 
{\cal L}^{(4)} &=& L_1 \langle D_\mu U^\dagger D^\mu U \rangle^2 +
  L_2 \langle D_\mu U^\dagger D_\nu U \rangle^2 \nonumber \\ &+& L_3
  \langle \left( D_\mu U^\dagger D^\mu U \right)^2\rangle \nonumber \\
  &+& L_4 \langle D_\mu U^\dagger D^\mu U \rangle \langle \chi^\dagger
  U + U^\dagger \chi \rangle \nonumber \\ &+& L_5 \langle D_\mu
  U^\dagger D^\mu U ( \chi^\dagger U + U^\dagger \chi) \rangle
  \nonumber \\ &+& L_6 \langle \chi^\dagger U + U^\dagger \chi
  \rangle^2 \nonumber \\ &+& L_7 \langle \chi^\dagger U - U^\dagger
  \chi \rangle^2 + L_8 \langle ( \chi^\dagger U)^2 + (U^\dagger
  \chi)^2 \rangle \nonumber \\ &-& iL_9 \langle F_{\mu\nu}^L D^\mu U
  D^\nu U^\dagger + F_{\mu\nu}^R D^\mu U^\dagger D^\nu U \rangle
  \nonumber \\ &+& L_{10} \langle F_{\mu\nu}^L U F^{\mu\nu}{}^R
  U^\dagger \rangle \nonumber \\ &+& H_1 \langle (F_{\mu\nu}^R)^2 +
  (F_{\mu\nu}^L)^2 \rangle + H_2 \langle \chi^\dagger \chi \rangle \,.
\label{eq:chl4_flat}
\end{eqnarray} 
Here, we have introduced the standard chiral covariant derivatives and
gauge field strength tensors,
\begin{eqnarray} 
D_\mu U &=& D_\mu^L U-U D_\mu^R =
\partial_\mu U-i A_\mu^L U +iU  A_\mu^R, \\ 
 F_{\mu\nu}^r &=& i[ D_\mu^r, D_\nu^r] = 
\partial_\mu A_\nu^r -\partial_\nu A_\mu^r
-i [ A_\mu^r , A_\nu^r ], \nonumber 
\end{eqnarray} 
with $r=L, R$. $ f$ is the pion weak decay constant in the chiral
limit and $ U = e^{ { i} \sqrt{2} \Pi / f } $ a unitary matrix and
$\Pi$ a non-linear transforming field under the chiral group
representing the octet of pseudoscalar mesons given by
\begin{eqnarray} 
        \Pi = \left( \matrix{ \frac{1}{\sqrt{2}} \pi^0 +
        \frac{1}{\sqrt{6}} \eta & \pi^+ & K^+  \cr  \pi^- & -
        \frac{1}{\sqrt{2}} \pi^0 + \frac{1}{\sqrt{6}} \eta & K^0  \cr 
        K^- & \bar{K}^0 & - \frac{2}{\sqrt{6}} \eta }
        \right) \,.
\label{eq:Pi}
\end{eqnarray}
Finally, $ \chi = 2 B_0 \,\text{diag}(m_u,m_d,m_s) $ is the current quark mass
matrix with $B_0 = |\langle \bar q q \rangle | / f^2 $ where $\langle
\bar q q \rangle $ is the quark condensate for one flavor.

Note that the coefficients $L_1-L_{10} $ appear in ${\cal L}^{(4)}$
given by Eq.~(\ref{eq:chl4_flat}).  As one sees, the terms containing
$L_{11}-L_{13}$ cannot be obtained by computing the energy momentum
tensor from the chiral effective Lagrangian in flat-space time and
from this viewpoint are genuine new contributions to $\theta^{\mu \nu}$.
The gravitational low energy coefficients arise at the hadronic level
due to quantum effects and corresponds to the inclusion of curvature
terms in the Lagrangian which generate purely transverse terms in the
energy momentum tensor. Actually, from a calculational viewpoint it is
more advantageous to couple the system to gravity and compute the
chiral Lagrangian in curved space-time.

Although the LEC's should be deduced from QCD itself solely in terms
of $\Lambda_{\rm QCD}$ and the quark masses, their numerical values
are obtained from a fit to low energy data, and systematic
calculations of the corresponding non-gravitational low-energy
constants (LEC's) $L_{1-10} $ have been carried out in the recent past
up to two loop
accuracy~\cite{Bijnens:1998fm,Colangelo:2001df,Amoros:2001cp,Bijnens:2002hp},
or by using the Roy equations~\cite{Ananthanarayan:2000ht} (see also
~\cite{Yndurain:2002ud,Pelaez:2003eh}). For strong and electroweak
processes involving pseudoscalar mesons the bulk of the LEC's is
saturated in terms of resonance
exchanges~\cite{Ecker:1988te,Polyakov:1995vh,Vereshagin:1996xa,Vereshagin:1998mg},
which can be justified in the large-$N_c$ limit in a certain
low-energy approximation~\cite{Pich:2002xy} by imposing relevant QCD
short-distance constraints. In the case of gravitational processes
similar ideas apply~\cite{Donoghue:1991qv}, although less model
independent information is known. (A one loop calculation has been
carried out in Ref.~\cite{Kubis:1999db}).

In a more model dependent set up the LEC's can also be understood from
a more microscopic viewpoint by using chiral quark
models~\cite{Diakonov:1984tw,Balog:1985ps,Andrianov:1985ay,
Belkov:1987mb,Hansson:1990jy,Espriu:1989ff,Praszalowicz:1989dh,
Holdom:1990iq,RuizArriola:1991gc,Bernard:1991wy,
Schuren:1991sc,Bijnens:1992uz,Schuren:1993aj,Wang:1999cp}
for the non-gravitational coefficients $L_1- L_{10}$. The calculation
of $L_{11},L_{12} $ and $L_{13}$, encoding the coupling to
gravitational sources, has only been done for a chiral quark model in
Ref.~\cite{Andrianov:1998fr} and more recently in the spectral quark
model~\cite{Megias:2004uj}. These values could be used to predict the
experimentally unmeasured gravitational form factors of pseudoscalar
mesons (see Ref.~\cite{Kubis:1999db} for a calculation within ChPT).

In this paper we provide the values of the gravitational LEC's
$L_{11,12,13}$, as well as $H_{0,3-5}$, in several chiral quark
models. Namely, we use in Sect.~\ref{sec:cqm} the Nambu--Jona-Lasinio
(NJL) model~\cite{Nambu:1961tp,Nambu:1961fr} and the Georgi-Manohar (GM)
model~\cite{Manohar:1983md} in the presence of gravity.  Although
these models look very different in appearance they do incorporate
chiral symmetry breaking and quarks have a constituent mass $M \sim
300{\rm MeV}$.  The main difference between them is that while in the
NJL model this mass is dynamically generated through interactions, the
GM already assumes that this breaking has taken place. Moreover, while
in the GM model there are only pseudoscalar Goldstone bosons as
dynamical degrees of freedom in the NJL there are additional $\bar q q
$ scalar fields.  Finally, in the NJL model quarks appear to have an
axial coupling constant $g_A $ for the quarks equal to one, while in
the GM model one starts already with $ g_A \neq 1$. A more detailed
analysis~\cite{RuizArriola:1991gc} shows that if in NJL $G_V \neq 0 $ then
$g_A \neq 1 $ (see also below).  More specifically, we consider the
generalized SU(3) NJL with scalar, pseudoscalar, vector and axial
couplings extending the original calculation of
Ref.~\cite{RuizArriola:1991gc} (see also Ref.~\cite{Bijnens:1992uz} where
also an estimate on the gluonic effects of the LEC's was done.). To
this end we apply, after bosonization and Pauli-Villars
regularization, a coordinate and frame invariant derivative expansion
by means of the heat kernel technique in curved space-time (Sect.~
\ref{sec:hk}). After that, the scalars, vectors and axial vectors are
eliminated at the mean field level (Sect.~\ref{sec:mf}) and the full
results are presented in Sect.~\ref{sec:res}.

For notation details, we rely heavily on previous
work~\cite{RuizArriola:1991gc,RuizArriola:1995ea,Megias:2004uj}. Nevertheless,
the inclusion of gravity is a bit messy, so we present the minimal
amount of formulas to make the paper more comprehensive and also to
point out some differences with previous calculations
non-gravitational LEC's by one of us (ERA)~\cite{RuizArriola:1991gc} and
other authors~\cite{Bijnens:1992uz} as well as our previous work with
Broniowski on the gravitational LEC's in the spectral quark
model~\cite{Megias:2004uj}.

\section{Chiral  quark  models in curved space time}
\label{sec:cqm}

The coupling of gravity\footnote{ We consider only Einstein gravity,
i.e., we use the Riemann connection, uniquely defined by being
torsionless and metric preserving (metricity). Extension to torsion
gravity \cite{Hammond:2002rm} is possible but will not be studied
here.} to chiral quark models has been described at length in
Ref.~\cite{Megias:2004uj} where more details on the notation and this
specific model can be found in the context of the Spectral Quark Model
(SQM)~\cite{RuizArriola:2001rr,RuizArriola:2003bs,RuizArriola:2003wi}.
In this paper we discuss several constituent chiral quark models which
have the common feature of incorporating dynamical chiral symmetry
breaking at the one quark loop level: The Nambu--Jona-Lasinio (NJL)
model~\cite{Nambu:1961tp,Nambu:1961fr} and the Georgi-Manohar (GM)
model~\cite{Manohar:1983md}. In these models quarks have a constituent
mass $M \sim 300{\rm MeV}$.  The main difference between them has to
do with the presence or not of dynamical $\bar q q $ scalar fields
respectively. In addition, while the NJL model dynamically generates
spontaneous chiral symmetry breaking, the GM model already starts in a
chirally broken phase. To avoid unnecessary duplication we will
discuss explicitly the NJL model with scalar and vector couplings from
which the relevant implications for the GM model can be more easily
deduced.

\subsection{Nambu--Jona-Lasinio model} 
\label{sec:njl} 

The Nambu--Jona-Lasinio (NJL) action in curved Minkowski space-time
with metric tensor $g_{\mu \nu} (x) $ reads
\begin{eqnarray}
S = \int d^4 x \sqrt{-g} {\cal L}_{\rm NJL} 
\end{eqnarray} 
where $ g = \Det( g_{\mu\nu} ) $ and the Lagrangian is given by 
\begin{eqnarray} 
{\cal L}_{\rm NJL} &=& \bar{q} (i\slashchar\partial + \slashchar\omega
- \hat{M}_0 )q \nonumber \\ &+& {G_S \over 2}\sum_{a=0}^{N_f^2-1}
\left( (\bar{q}\lambda_a q)^2 +(\bar{q}\lambda_a i \gamma_5 q)^2
\right) \nonumber \\ &-& {G_V \over 2}\sum_{a=0}^{N_f^2-1} \left(
(\bar{q}\lambda_a \gamma_\mu q)^2 + (\bar{q}\lambda_a \gamma_\mu
\gamma_5 q )^2 \right)
\end{eqnarray} 
where $q=(u,d,s, \dots )$ represents a quark spinor with $N_c $
colors and $N_f$ flavors. The $\lambda$'s represent the Gell-Mann
matrices of the $U(N_f)$ group and $\hat M_0= {\rm diag} (m_u, m_d,
m_s,\ldots) $ stands for the current quark mass matrix.  In the limiting
case of vanishing current quark masses the classical NJL-action is
invariant under the global $U(N_f)_R \otimes U(N_f)_L $ group of
transformations. The derivative $\partial_\mu - i \omega_\mu $ is
frame (local Lorentz) and general-coordinate covariant and it includes
the spin connection,
\begin{eqnarray} 
\omega_\mu (x) = \frac{i}8 \left[ \gamma^\nu (x) ,
\gamma_{\nu;\mu} (x) \right] 
\end{eqnarray} 
where the frame covariant derivative is defined as usual 
\begin{eqnarray} 
\gamma_{\nu  ;\mu}=  \partial_\mu 
\gamma_\nu (x) - \Gamma_{\nu \mu}^\lambda (x) \gamma_\lambda (x)
\end{eqnarray} 
and $ \Gamma_{\lambda \mu}^\sigma $ are the Christoffel symbols which
for a Riemannian space read,
\begin{eqnarray}
\Gamma_{\lambda \mu}^\sigma = \frac{1}{2} g^{\nu \sigma} \left\{
\partial_\lambda g_{\mu\nu} + \partial_\mu g_{\lambda \nu} -
\partial_\nu g_{\mu \lambda} \right\} .
\label{eq:christoffel}
\end{eqnarray} 
The space-time dependent Dirac matrices, $\gamma_\mu (x) $, fulfill
the standard anticommutation rules $ \{ \gamma^\mu (x) , \gamma^\nu
(x) \} = 2 g^{\mu \nu} (x) $. The $\gamma_5 $ matrix is, on the
contrary, $x$-independent (we use the conventions of
Ref.~\cite{Itzykson:1980bk}). The Dirac slash is constructed as $ i
\slashchar\partial q (x) = i \gamma^\mu (x) \partial_\mu q(x)
$~\footnote{Note that in the curved case $\partial^\mu \varphi (x) =
g^{\mu \nu} (x) \partial_\nu \varphi (x) $, and the self-adjoint
D'Alambertian on a spin-0 field is given by $ \partial^2 \varphi (x) =
1/\sqrt{-g} \partial_\mu ( \sqrt{-g} g^{\mu \nu} \partial_\nu \varphi
(x) ) $ so that
$$
 \int d^4 x \sqrt{-g} \varphi_1(x) \partial^2 \varphi_2 (x) = \int d^4
x \sqrt{-g} \varphi_2(x) \partial^2 \varphi_1 (x)  \,.
$$
Likewise, for a Dirac spinor the form 
$\int d^4x\sqrt{-g}\bar\psi_1( i\slashchar\partial +
\slashchar\omega)\psi_2 $ is Hermitian.}. 

The vacuum to vacuum transition amplitude in the presence of external
scalar, pseudoscalar, vector, axial and metric bosonic fields
$(s,p,v,a,g)$ Lagrangian can be written as a path integral in the form
\begin{eqnarray}
Z[s,p,v,a,g] &=& \langle 0 | {\rm T} \exp \Bigl\{ i \int d^4 x \sqrt{-g}
 {\cal L}_{\rm ext} (x)  \Bigr\} |0 \rangle \nonumber \\ &=&
\int D\bar{q} Dq \exp \Bigl\{i\int d^4 x \sqrt{-g} \left[{\cal L}_{\rm NJL} 
+ {\cal L}_{\rm ext} \right] \Bigr\}  \,, \nonumber \\ 
\end{eqnarray} 
where the external field Lagrangian is given by
\begin{eqnarray}
{\cal L}_{\rm ext} (x) = \bar q \({\slashchar v}+{\slashchar a}
\gamma_5 -(s+i\gamma_5 p)\)q \,.
\end{eqnarray} 
Following the standard procedure~\cite{Eguchi:1976iz} it is convenient to
introduce auxiliary bosonic fields $(S,P,V,A)$ so that after formally
integrating the quarks out one gets the equivalent generating
functional  
\begin{eqnarray} 
Z[s,p,v,a,g] &=& \int DS DP DV DA e^{ i \Gamma[{S,P,V,A;s,p,v,a,g]}}
\nonumber \\
\end{eqnarray} 
 where the effective action 
\begin{eqnarray}
\Gamma[S,P,V,A;s,p,v,a,g] &=&-{\rm i} N_c {\rm Tr} \log \left( i {\bf
D} \right) \nonumber \\ 
&+& \int d^4 x \sqrt{-g} 
{\cal L}_m 
\label{eq:eff_ac} 
\end{eqnarray} 
have been introduced~\footnote{For a bilocal (Dirac and flavor
matrix valued) operator $A(x,x')$ one has 
$$
{\rm Tr} A = \int d^4 x
\sqrt{-g} \,{\rm tr} \langle A(x,x) \rangle \, , 
$$
with ${\rm tr} $ denoting the Dirac trace and $\langle \, \rangle $
the flavor trace.}. Here, the Dirac
operator is given by
\begin{eqnarray}
i {\bf D} &=& i\slashchar{\partial} + \slashchar{\omega} - {\hat M_0} + \left(
\slashchar{\V} + \slashchar{\A} \gamma_5 - \oS - i \gamma^5 \oP \right) \nonumber
\label{eq:dirac_op} 
\end{eqnarray} 
where we introduce the additive combinations 
\begin{eqnarray}
\overline{S} = s + S , \, \quad  
\overline{P} = p + P , \, \quad  
\overline{V} = v + V , \, \quad  
\overline{A} = a + A \nonumber \\  
\end{eqnarray} 
and the mass term 
\begin{eqnarray} 
{\cal L}_m &=& -{1\over 4G_S} \langle S^2
+ P^2 \rangle  + {1\over 4G_V} \langle  V_\mu V^\mu + A_\mu A^\mu \rangle 
\end{eqnarray} 
has been defined. Here, $(S,P,V,A)$ are dynamical scalar,
pseudoscalar, vector and axial meson fields respectively, whereas
$(s,p,v,a,g)$ represent external sources.  With the exception of the
metric tensor $g_{\mu \nu} (x)$, all fields are expanded in terms of
the $\lambda$ flavor matrices. Notice also that for the path integral
in the bosonic fields to be well defined in Minkowski space we must
use the prescription $G_S^{-1} \to G_S^{-1} -i \epsilon $ and $
G_V^{-1} \to G_V^{-1} -i\epsilon $.
\vspace{0.1cm}

The additive contribution to the effective action can be separated
into a $\gamma_5$-even and $\gamma_5$-odd parts corresponding to
normal and abnormal parity processes respectively. Introducing
the operator~\footnote{$\D_5 $ corresponds to rotate $\D$ to Euclidean
space, take the hermitian conjugate and rotate back to Minkowski
space.}
\begin{eqnarray}
\D_5 [ \oS, \oP, \V, \A ] = \gamma_5 \D [ \oS,-\oP,\V,-\A] \gamma_5 \,,
\end{eqnarray} 
the $\gamma_5$-even contribution is quadratically divergent and can be
regularized in a chiral gauge invariant manner by means of the
Pauli-Villars scheme~\cite{Pauli:1949zm} 
\begin{eqnarray} 
\Gamma_{\rm even} &=& -i{N_c \over 2} {\rm Tr} \sum c_i \log(\D \D_5 +
\Lambda_i^2 +i\epsilon) \\ 
&& -{1\over 4G_S} \langle S^2 +
P^2 \rangle + {1\over 4G_V} \langle V_\mu V^\mu + A_\mu A^\mu \rangle
\nonumber 
\end{eqnarray} 
where the Pauli-Villars regulators fulfill $c_0=1$, $\Lambda_0 =0$ and
the conditions $\sum_i c_i =0$, $\sum_i c_i \Lambda_i^2 =0 $ which
render finite the logarithmic and quadratic divergences
respectively. In practice, we take two cut-offs in the coincidence
limit $ \Lambda_1 \to \Lambda_2 = \Lambda $ and hence $ \sum_i c_i
f(\Lambda_i^2)=f(0)-f(\Lambda^2)+\Lambda^2 f^\prime(\Lambda^2) $. In
the vacuum ($s=p=v=a=0$) one has spontaneous chiral symmetry breaking,
since by minimizing $\Gamma_{\rm even}$ one gets $S=M$ with $M$ the
constituent quark mass and $P=V=A=0$, and the gap equation
\begin{eqnarray}
\frac{1}{G_S}  &=& -i 4 N_c \sum_i c_i \int \frac{d^4 k}{(2\pi)^4}
\frac{1}{k^2-M^2-\Lambda_i^2}
\label{eq:gap}
\end{eqnarray} 
where the Pauli-Villars regularization has been used. Thus, according
to Goldstone's theorem we will write the decomposition
\begin{equation}
{\cal M} \equiv \oS+ i \oP = M U + \overline{m}  
\end{equation} 
with $ U = e^{ { i} \sqrt{2} \Pi / f } $ and $\Pi$ is given by
Eq.~(\ref{eq:Pi}). 

\subsection{Georgi-Manohar model} 
\label{sec:gm} 

In the presence of gravity, the Georgi-Manohar model
Lagrangian~\cite{Manohar:1983md} reads
\begin{eqnarray} 
{\cal L}_{\rm GM} &=& \bar{q} \Bigg( i\slashchar\partial +
\slashchar\omega - M U^5 - \hat{M}_0  \nonumber \\ 
&+& \frac{1}{2}(1-g_A) U^5 i\slashchar\partial
U^5 \Bigg) q
\label{eq:GM} 
\end{eqnarray} 
where $U^5 = e^{ { i} (\sqrt{2} \Pi / f) \gamma_5 } $, and $ \Pi $ is
given by Eq.~(\ref{eq:Pi}), $M$ the constituent quark mass, and $ g_A$
the axial coupling of the quarks, which will be assumed as suggested
in Ref.~\cite{Manohar:1983md} to be different from one. The generating
functional now reads,
\begin{eqnarray} 
Z[s,p,v,a,g] &=& \int DU  e^{ i \Gamma[U;s,p,v,a,g]}
\nonumber \\
\end{eqnarray} 
with the effective action given by 
\begin{eqnarray}
\Gamma[U;s,p,v,a,g] &=&-{\rm i} N_c {\rm Tr} \log \left( i {\bf
D} \right)  \,.
\label{eq:eff_ac_gm} 
\end{eqnarray} 
Direct comparison with the one quark loop effective action of the NJL,
Eq.~(\ref{eq:eff_ac}) shows that it corresponds to a similar model
without mass terms ${\cal L }_m $ and with a Dirac operator as in
Eq.~(\ref{eq:dirac_op}) with a specific choice of the spin-1 dynamical field 
operators, 
\begin{eqnarray} 
V_\mu &=& \frac{1}{4}(1- g_A) \left[ U^\dagger \partial_\mu U - \partial_\mu
U^\dagger U \right] \,,\label{eq:va_gm1} \\
A_\mu &=& \frac{1}{4}(1- g_A) \left[ U^\dagger \partial_\mu U + \partial_\mu
U^\dagger U \right] \,. 
\label{eq:va_gm2}
\end{eqnarray} 
In Eq.~(\ref{eq:eff_ac_gm}) we understand that the same Pauli-Villars
regularization as in the NJL model has been implemented.

\section{Heat kernel expansion} 
\label{sec:hk}

In the chiral expansion of the action, Eq.~(\ref{eq:eff_ac}), the
counting is the standard one, the pseudoscalar field $U$ and the
curved space-time metric $g_{\mu \nu} $ are zeroth order, the vector
and axial fields $ v_\mu $ and $a_\mu$ are first order, and any
derivative $\partial_\mu $ is taken to be first order. The external
scalar and pseudoscalar fields $s$ and $p$ and the current mass matrix
$\hat m_0$ are taken to be second order. At the one quark loop level
this chiral expansion corresponds to a derivative expansion which
should be gauge, frame and coordinate invariant.

In order to carry out the low energy expansion of the action via a
heat kernel method it proves useful to use the proper-time integral
representation of the logarithm
\begin{eqnarray}
\sum_i c_i {\rm Tr} \log \left( {\bf D}_5 {\bf D} + \Lambda_i^2
\right) = - {\rm Tr} \int_0^\infty \frac{d \tau}{\tau} e^{- {\rm i}
\tau {\bf D}_5 {\bf D} } \phi(\tau)   
\end{eqnarray} 
where $\phi (\tau) $ is the proper time representation of the
Pauli-Villars regularization, 
\begin{eqnarray}
\phi(\tau) = \sum_i c_i e^{-i \tau \Lambda_i^2} 
\end{eqnarray}
with the conditions 
\begin{eqnarray}
\sum_i c_i =0 \qquad \sum_i c_i \Lambda_i^2 =0  
\end{eqnarray}
which fulfills the conditions $ \phi(0)=0 $ and $ \phi'(0)=0$, thus
killing the quadratic and logarithmic divergencies respectively.  The
operator inside the logarithm is of Klein-Gordon type in curved
space-time with some spinorial structure,
\begin{eqnarray}
{\bf D}_5 {\bf D} &=& \frac{1}{\sqrt{-g}} \left[ {\cal D}_\mu \left( 
\sqrt{-g} g^{\mu \nu} {\cal D}_\nu \right) \right] + {\cal V},
\label{eq:KG-heat} 
\end{eqnarray} 
with 
\begin{eqnarray}
{\cal V}=  {\cal V}_R P_R + {\cal V}_L  P_L   
\end{eqnarray} 
and 
\begin{eqnarray}
{\cal V}_R &=& -\frac{1}{2} \sigma^{\mu \nu} {F}_{\mu \nu}^R  
+ \frac{1}{4} R - {\rm i} \gamma^\mu {\nabla}_\mu  {\cal M} + {\cal
M}^\dagger {\cal M} ,\nonumber \\ \\ 
{\cal V}_L &=& -\frac{1}{2} \sigma^{\mu \nu} {F}_{\mu \nu}^L  
+ \frac{1}{4} R - {\rm i} \gamma^\mu {\nabla}_\mu  {\cal M}^\dagger + {\cal
M} {\cal M}^\dagger \,.   \nonumber 
\end{eqnarray} 
As we saw in the previous section, the one quark loop contribution to
the effective action depends additively on the external and dynamical
fields. Although this complicates matters one can borrow from previous
results taking into account some minor modifications regarding this
fact and the regularization~\footnote{This additivity works only for
the normal parity contribution of the action, but must be
modified in the abnormal parity sector in order to reproduce
the QCD anomaly. See e.g. Ref.~\cite{Bijnens:1993cy,RuizArriola:1995ea}}.

The form of the operator $ {\bf D}_5 {\bf D} $ in
Eq.~(\ref{eq:KG-heat}) is suitable to make a heat kernel expansion in
curved space-time as the one of Ref.~\cite{Luscher:1982wf}. For a
review see e.g. \cite{Vassilevich:2003xt} and references therein. In
our particular case, before undertaking the heat kernel expansion we
separate a constituent quark mass squared, $M^2$, contribution from
the operator ${\bf D}_5 {\bf D}$ which we treat exactly,
\begin{eqnarray}
\langle x | e^{-{\rm i} \tau {\bf D}_5 {\bf D} } | x \rangle &=& e^{ -
{\rm i}\tau M^2 } \langle x | e^{-{\rm i} \tau ( {\bf D}_5 {\bf D} -
M^2 ) } | x \rangle \\ &=& \frac{\rm i}{(4\pi {\rm i}\tau )^2}
e^{-{\rm i} \tau M^2 } \sum_{n=0}^\infty a_{n} (x) \left( {\rm i}
\tau \right)^n \,. \nonumber
\end{eqnarray}
The $\tau$ integrals appearing in the effective action are of the form
\begin{equation}
\I_{2l} := M^{2l} \int_0^\infty \frac{d\tau}{\tau}\phi(\tau)(i\tau)^l
e^{-i\tau M^2} \,.
\end{equation} 
For completeness, we list here the particular values,
\begin{eqnarray}
M^4 \I_{-4} &=& -\frac{1}{2}\sum_i
c_i(\Lambda_i^2+M^2)^2\log(\Lambda_i^2+M^2)
\,, \\ 
M^2\I_{-2} &=& \sum_i
c_i(\Lambda_i^2+M^2)\log(\Lambda_i^2+M^2) \,, \\ \I_0 &=& -\sum_i
c_i\log(\Lambda_i^2+M^2) \,, \\ \I_{2n} &=&\Gamma(n)\sum_i c_i
\left(\frac{M^2}{\Lambda_i^2+M^2}\right)^n ,
\textrm{Re}(n)>0 \,.
\end{eqnarray}

After evaluation of the Dirac traces, the second order Lagrangian is given by 
\begin{eqnarray}
{\cal L}^{(2)}_q  &=&{N_c\over(4\pi)^2} \Big\{ M^2 {\cal I}_0 \langle
\Nabla_\mu U^\dagger \Nabla^\mu U \rangle \\ &+& 2 M^3 {\cal
I}_{-2}\langle \overline{\it m}^\dagger U + U^\dagger \overline{\it m} \rangle
+ \frac{M^2}{6}{\cal I}_{-2} \langle R \rangle \Big\}, \nonumber 
\end{eqnarray} 
whereas the fourth order becomes
\begin{eqnarray} 
{\cal L}^{(4)}_q  &=& {N_c\over (4\pi)^2} \Big\{  
\nonumber \\
&-& {1\over6} {\cal I}_0  \langle (\F_{\mu\nu}^R)^2 +
(\F_{\mu\nu}^L)^2 \rangle \nonumber \\ &+& {\cal I}_0 \langle
\frac{7}{720} R_{\alpha \beta \mu \nu} R^{\alpha \beta \mu \nu} - 
\frac{1}{144} R^2 + \frac{1}{90} R_{\mu \nu} R^{\mu \nu} \rangle
\nonumber \\
&-&\frac{i }2 {\cal I}_2 \langle \F_{\mu\nu}^R \Nabla^\mu U^\dagger
\Nabla^\nu U + \F_{\mu\nu}^L \Nabla^\mu U \Nabla^\nu U^\dagger \rangle
\nonumber \\
&+& \frac{1}{12} {\cal I}_4 \langle (\Nabla_\mu U \Nabla_\nu U^\dagger
)^2 \rangle -\frac{1}{6} {\cal I}_4 \langle (\Nabla_\mu U \Nabla^\mu
U^\dagger )^2 \rangle \nonumber \\
&+& \frac{1}{6} {\cal I}_2 \langle \Nabla_\mu \Nabla_\mu U \Nabla^\nu \Nabla^\nu 
U^\dagger \rangle \nonumber \\ 
&+&  2 M^2{\cal I}_{-2} \langle \overline{\it m}^\dagger\overline{\it m} \rangle - M^2 {\cal I}_0   \langle ( \overline{\it m}^\dagger
U+U^\dagger \overline{\it m} )^2 \rangle  
\nonumber \\
&-& M {\cal I}_2 \langle \Nabla_\mu U^\dagger \Nabla^\mu U
(\overline{\it m}^\dagger U+U^\dagger \overline{\it m} ) \rangle \nonumber \\
&+& M {\cal I}_0 \langle \Nabla_\mu U^\dagger \Nabla^\mu \overline{\it m} +
\Nabla_\mu \overline{\it m}^\dagger \Nabla^\mu U \rangle \nonumber \\ 
&-& \frac{M}{6}{\cal I}_0 R\langle U^\dagger \overline{\it m}+
\overline{\it m}^\dagger U \rangle - \frac{1}{12} {\cal I}_2 R\,
\langle \Nabla_\mu U^\dagger \Nabla^\mu U \rangle \Big\} \,. \nonumber \\
\label{eq:L4}
\end{eqnarray} 
Here $R^\lambda_{\sigma \mu \nu}$, $R_{\mu \nu} $ and $R$ are the
Riemann curvature tensor, the Ricci tensor, and the curvature scalar,
respectively\footnote{Note the opposite sign of our definition for
the Riemann tensor as compared to Ref.~\cite{Donoghue:1991qv}. We follow
Ref.~\cite{Weinberg:1972bk}.},
\begin{eqnarray}
-R^\lambda_{\,\, \sigma \mu \nu} &=& \partial_\mu \Gamma^\lambda_{\nu \sigma}
- \partial_\nu \Gamma^\lambda_{\mu \sigma}+ \Gamma^\lambda_{\mu
\alpha} \Gamma^\alpha_{\nu \sigma} - \Gamma^\lambda_{\nu \alpha}
\Gamma^\alpha_{\mu \sigma}, \nonumber \\ 
R_{\mu \nu} &=& R^\lambda_{\, \, \mu \lambda \nu } \, ; 
\quad R = g^{\mu \nu} R_{\mu
\nu} \, .
\label{eq:curvature}    
\end{eqnarray}
The Riemann connection is given by the Christoffel symbols, $
\Gamma_{\lambda \mu}^\sigma $ defined in Eq.~(\ref{eq:christoffel}).

We have introduced the gauge and frame covariant derivatives and
the field strength tensor containing the external and internal
(bosonized) degrees of freedom, 
\begin{eqnarray} 
\Nabla_\mu U &=& \nabla_\mu U -i V_\mu^L U +iU  V_\mu^R, \\ 
\overline{F}_{\mu\nu}^r &=& 
\partial_\mu \A_\nu^r -\partial_\nu \A_\mu^r
-i [ \A_\mu^r , \A_\nu^r ], \nonumber 
\end{eqnarray} 
with $r=L, R$ and the spin 0 additive combination 
\begin{eqnarray}
\overline{m} &=& (S+ i P- M U ) + m  \,.
\end{eqnarray} 
The form of Eq.~(\ref{eq:L4}) is not yet ready for comparison with the
result of Ref.~\cite{Gasser:1984gg,Donoghue:1991qv}. To do that we have
to eliminate all degrees of freedom other than the on-shell
pions.  We proceed in three steps: Firstly we integrate out the vector
and axial degrees of freedom, afterwards we eliminate the scalars and
finally we exploit the classical equations of motion for the
pseudoscalars.

\section{Mean field Equations of Motion} 
\label{sec:mf} 

\subsection{Elimination of Vector and Axial Vectors} 
\label{sec:mf_va}

To eliminate the $V_\mu $ and $ A_\mu$ fields at the mean field level
we minimize the corresponding Lagrangian with respect to those
fields. To the order in the chiral expansion we are working it is
enough to deal with the terms containing vector mesons with two
Lorentz indices, i.e. the mass term and the second order term stemming
from the quark determinant,
\begin{eqnarray}
{\cal L}^{(2)}_{A,V} &=&{N_c\over(4\pi)^2} M^2 {\cal I}_0 \langle
\Nabla_\mu U^\dagger \Nabla^\mu U \rangle + {1\over 4G_V} \langle
V_\mu V^\mu + A_\mu A^\mu \rangle \,. \nonumber \\
\end{eqnarray}
The result looks like the GM model
Eqs.~(\ref{eq:va_gm1},\ref{eq:va_gm2}) as noted in
Ref.~\cite{RuizArriola:1991gc},
\begin{eqnarray}
\V_\mu^R &=& v_\mu^R +
\frac{i}{2}(1-g_A) U^\dagger\nabla_\mu U \,, 
\\
\V_\mu^L &=& v_\mu^L + \frac{i}{2}(1-g_A) U \nabla_\mu
U^\dagger \,, 
\end{eqnarray} 
which looks like a chiral gauge transformation of the additive fields
$\V$ and $\A$. Then it is straightforward to obtain the following
relations,
\begin{eqnarray}
\F_{\mu\nu}^R &=& \frac{1}{2}(1+g_A)
F_{\mu\nu}^R+\frac{1}{2}(1-g_A)U^\dagger F_{\mu\nu}^L
U \nonumber \\ &-& \frac{i}{4}(1-g_A^2) \left(\nabla_\mu U^\dagger
\nabla_\nu U -\nabla_\nu U^\dagger \nabla_\mu U\right) \,,
\\
\F_{\mu\nu}^L &=& \frac{1}{2}(1-g_A) U F_{\mu\nu}^R
U^\dagger+\frac{1}{2}(1+g_A) F_{\mu\nu}^L \nonumber \\ 
&-& \frac{i}{4}(1-g_A^2) \left(\nabla_\mu U
\nabla_\nu U^\dagger -\nabla_\nu U \nabla_\mu U^\dagger\right) \,, 
\\
\Nabla_\mu U &=& g_A \nabla_\mu U \, , \\ 
\Nabla^2 U &=& g_A \nabla^2 U + ig_A (1-g_A) U \nabla_\mu U^\dagger
\nabla^\mu U \,. \label{eq:umu}
\end{eqnarray}

\subsection{Elimination of Scalars} 
\label{sec:mf_s}

The elimination of scalars proceeds along similar lines as in the
vector and axial case. If we do a chiral rotation 
\begin{eqnarray}
S+ iP = \sqrt{U} \Sigma \sqrt{U}   
\end{eqnarray} 
with $ \Sigma^\dagger = \Sigma $, and using that $\Sigma = M + \Phi $
with $\Phi$ the fluctuation around the vacuum value we have
\begin{eqnarray}
{\overline m} &=&  \sqrt{U} \Phi \sqrt{U} + \frac{1}{2B_0} \chi  \,.
\end{eqnarray}
The mass term becomes 
\begin{eqnarray} 
{\cal L}_m &=& -{1\over 4G_S} \langle M^2 + 2 M \Phi + \Phi^2 \rangle \,.
\end{eqnarray} 
Using the gap equation, Eq.~(\ref{eq:gap}), linear terms in $\Phi$ not
containing external fields vanish. As a consequence, the part of the
Lagrangian comprising the scalar field is given by
\begin{eqnarray}
 \cL_\Phi (x) &=& -\frac{N_c}{(4\pi)^2} \Bigg\langle 4 M^2 {\cal I}_0
   \Phi^2 + \frac{1}{3} M {\cal I}_0 R \Phi \nonumber \\ 
   &+& M {\cal I}_0
   \sqrt{U} \Phi \sqrt{U^\dagger } \left( U \Nabla^2 U^\dagger +
   \Nabla^2 U U^\dagger \right) \nonumber \\  
   &+& 2M^2(2 {\cal I}_0 - {\cal I}_{-2} ) \sqrt{U} \Phi \sqrt{U^\dagger }
   ( U m^\dagger + U^\dagger m ) \nonumber \\ 
   &+& 2 M {\cal I}_2 \sqrt{U}
   \Phi \sqrt{U^\dagger } \Nabla_\mu U \Nabla^\mu U^\dagger \Bigg\rangle 
\end{eqnarray}
from which the well known mass formula $ M_S = 2 M $ follows.
Minimizing with respect to the field $\Phi $ the classical equation of
motion follow,
\begin{eqnarray}
\sqrt{U} \Phi \sqrt{U^\dagger } &=& -\frac{1}{24M} R + \frac{1}{4M} \left( 1- \frac{{\cal I}_2}{{\cal
I}_0} \right) \Nabla_\mu U \Nabla^\mu U^\dagger \nonumber \\ &-&
\frac{1}{2}\left( 1 -\frac{{\cal I}_{-2}}{2 {\cal I}_0} \right) ( U {\it m}^\dagger + {\it m} U^\dagger) \,.
\end{eqnarray}
Substituting this equation into the Lagrangian $\cL_\Phi$ we obtain the
contribution to the effective Lagrangian stemming from the integration
of scalars.

\subsection{Mean field equations for pseudoscalars} 
\label{sec:mf_ps} 

The relevant equations of motion for the non-linear $U$ field are
obtained by minimizing ${\cal L}^{(2)}$. One obtains a set of
relations which are valid in the presence of curvature,
\begin{eqnarray} 
\langle \nabla^2 U^\dagger \nabla^2 U \rangle &=&  \langle \left(
\nabla_\mu U^\dagger \nabla^\mu U \right)^2 \rangle 
-\frac{1}{4} \langle \left(
\chi^\dagger U - U^\dagger \chi \right)^2 \rangle \nonumber \\ 
&+&\frac{1}{12} \langle 
\chi^\dagger U - U^\dagger \chi  \rangle^2 
\label{eq:id1}
\end{eqnarray} 
and 
\begin{eqnarray} 
\langle \chi^\dagger \nabla^2 U + \nabla^2 U^\dagger \chi \rangle &=& 2 \langle
\chi^\dagger \chi \rangle - \frac{1}{2} \langle \left( \chi^\dagger U +
U^\dagger \chi \right)^2 \rangle \nonumber \\ &-& \langle \left(
\chi^\dagger U + U^\dagger \chi \right) \nabla^\mu U^\dagger \nabla_\mu U
\rangle \nonumber \\ 
&+& \frac{1}{6} \langle \chi^\dagger U +
U^\dagger \chi  \rangle^2 \,.
\label{eq:id2}
\end{eqnarray} 
In the case of the $U(3)$ group one has $\Det \, U = e^{i \eta_0 /f}$
not necessarily equal to unity and the last two terms involving $
\langle \chi^\dagger U \pm U^\dagger \chi \rangle^2 $ in
Eqs.~(\ref{eq:id1}) and (\ref{eq:id2}) should be dropped. There is
another integral identity which proves very useful
\begin{eqnarray}
&& \int d^4 x \sqrt{-g} \, \langle \nabla_\mu \nabla_\nu U^\dagger
\nabla^\mu \nabla^\nu U \rangle = \int d^4 x \sqrt{-g} \langle
\nabla^2 U^\dagger \nabla^2 U \rangle \nonumber \\ 
&& i\int d^4 x \sqrt{-g} \, 
\langle F_{\mu\nu}^R \nabla^\mu U^\dagger \nabla^\nu U 
+F_{\mu\nu}^L \nabla^\mu U \nabla^\nu U^\dagger \rangle 
\nonumber  \\
&&-\int d^4 x \sqrt{-g} 
\langle F_{\mu\nu}^L U F^{\mu\nu}{}^R U^\dagger
\rangle
\nonumber \\
&& + \frac{1}{2}\int d^4 x \sqrt{-g}
 \langle (F_{\mu\nu}^R)^2 +(F_{\mu\nu}^L)^2  \rangle
\nonumber \\
&& + \int d^4 x
\sqrt{-g} R^{\mu \nu} \langle \nabla_\mu U^\dagger \nabla_\nu U
\rangle \,.
\label{eq:ricci}
\end{eqnarray} 
Finally, we also have the SU(3) identity
\begin{eqnarray}
\langle (\nabla_\mu U^\dagger \nabla_\nu U)^2 \rangle &=& - 2 \langle (
\nabla_\mu U^\dagger \nabla^\mu U )^2 \rangle \label{eq:su3} \\ &+& \langle \nabla_\mu
U^\dagger \nabla_\nu U \rangle^2 + \frac{1}{2} \langle \nabla_\mu
U^\dagger \nabla^\mu U \rangle^2 \,.
\nonumber 
\end{eqnarray}

\section{Results for the Gasser-Leutwyler--Donoghue coefficients} 
\label{sec:res}

Following the steps of Ref.~\cite{Megias:2004uj} one gets the chiral
effective Lagrangian in the presence of gravity in the form of Gasser,
Leutwyler and Donoghue form~\cite{Gasser:1984gg,Donoghue:1991qv}
\begin{eqnarray} 
{\cal L} &=& {\cal L}^{(0)}+{\cal L}^{(2,g)} + {\cal L}^{(2,R)} +
{\cal L}^{(4,g)} + {\cal L}^{(4,R)} + \dots
\label{eq:chl}
\end{eqnarray} 
%with the metric (upper-script $g$) and curvature (upper-script $R$)
%terms explicitly separated. 
The zeroth order vacuum contribution reads
\begin{eqnarray} 
{\cal L}^{(0)} = B = -\frac{2N_f N_c}{(4\pi)^2} M^4 {\cal I}_{-4} \, , 
\label{eq:chi0} 
\end{eqnarray} 
where $B$ is the vacuum energy density constant. The metric
contributions read
\begin{eqnarray} 
{\cal L}^{(2,g)}  &=& {f^2\over 4} \langle \nabla_\mu U^\dagger \nabla^\mu U
+(\chi^\dagger U + U^\dagger \chi) \rangle ,
\label{eq:chl2}
\end{eqnarray} 
and 
\begin{eqnarray} 
{\cal L}^{(4,g)} &=& L_1 \langle \nabla_\mu U^\dagger \nabla^\mu U
  \rangle^2 + L_2 \langle \nabla_\mu U^\dagger \nabla_\nu U \rangle^2
  \nonumber \\ &+& L_3 \langle \left( \nabla_\mu U^\dagger \nabla^\mu
  U \right)^2\rangle \nonumber \\ &+& L_4 \langle \nabla_\mu U^\dagger
  \nabla^\mu U \rangle \langle \chi^\dagger U + U^\dagger \chi \rangle
  \nonumber \\ &+& L_5 \langle \nabla_\mu U^\dagger \nabla^\mu U (
  \chi^\dagger U + U^\dagger \chi) \rangle \nonumber \\ &+& L_6
  \langle \chi^\dagger U + U^\dagger \chi \rangle^2 \nonumber \\ &+&
  L_7 \langle \chi^\dagger U - U^\dagger \chi \rangle^2 + L_8 \langle
  ( \chi^\dagger U)^2 + (U^\dagger \chi)^2 \rangle \nonumber \\ &-&
  iL_9 \langle F_{\mu\nu}^L \nabla^\mu U \nabla^\nu U^\dagger + F_{\mu\nu}^R
  \nabla^\mu U^\dagger \nabla^\nu U \rangle \nonumber \\ &+& L_{10} \langle
  F_{\mu\nu}^L U F^{\mu\nu}{}^R U^\dagger \rangle \nonumber \\ &+& H_1
  \langle (F_{\mu\nu}^R)^2 + (F_{\mu\nu}^L)^2 \rangle + H_2 \langle
  \chi^\dagger \chi \rangle \, .
\label{eq:chl4}
\end{eqnarray} 
%Here, we have introduced the standard chiral covariant derivatives and
%gauge field strength tensors,
%\begin{eqnarray} 
%\nabla_\mu U &=& \nabla_\mu^L U-U \nabla_\mu^R =
%\partial_\mu U-i A_\mu^L U +iU  A_\mu^R, \\ 
% F_{\mu\nu}^r &=& i[ \nabla_\mu^r, \nabla_\nu^r] = 
%\partial_\mu A_\nu^r -\partial_\nu A_\mu^r
%-i [ A_\mu^r , A_\nu^r ], \nonumber 
%\end{eqnarray} 
%with $r=L, R$. $ f$ is the pion weak decay constant in the chiral
%limit.  
The curvature contributions to the chiral Lagrangian can be
written in the form proposed in Ref.~\cite{Donoghue:1991qv} and are given
by
\begin{eqnarray}
{\cal L}^{(2,R)} &=& -H_0 R \, 
\end{eqnarray} 
and 
\begin{eqnarray}
{\cal L}^{(4,R)} &=& -L_{11} R \langle \nabla_\mu U^\dagger \nabla^\mu
U \rangle -L_{12} R^{\mu \nu} \langle \nabla_\mu U^\dagger \nabla_\nu
U \rangle \nonumber \\ &-& L_{13} R \langle \chi^\dagger U + U^\dagger
\chi \rangle + H_3 R^2 + H_4 R_{\mu \nu} R^{\mu \nu} \nonumber \\ &+&
H_5 R_{\mu \nu \alpha \beta} R^{\mu \nu \alpha \beta}.
\label{eq:chl4R}
\end{eqnarray} 
The curvature terms reflect the composite nature of the pseudoscalar
fields, since in the considered model they correspond to the coupling
of the gravitational external field at the quark level. 

Note that the pseudoscalar matrix $U$ is a coordinate and frame
scalar. So, once and only once the identities
(\ref{eq:id1})-(\ref{eq:su3}) have been used one can substitute
coordinate-frame covariant derivative by the covariant derivative,
i.e., $\nabla_\mu U = D_\mu U$. In this way Eqs.~(\ref{eq:chl2_flat})
and (\ref{eq:chl4_flat}) of this manuscript, and Eq. $(10)$ of
Ref.~\cite{Donoghue:1991qv} are deduced.

\subsection{GM model} 
\label{sec_gm_res} 

The Gasser-Leutwyler-Donoghue coefficients for the GM model do not
contain any contributions from the scalars, i.e. the spin $0$
field. So, the only contribution stems from the quark loop. For
this model, the pion weak decay constant is
\begin{equation}
f^2 = \frac{N_c}{4\pi^2}g_A^2 M^2 \I_0 \,.  \label{eq:f0GM}
\end{equation}
The normalization factor for the field $\chi$ is
\begin{equation}
B_0 = \frac{M}{g_A^2}\frac{\I_{-2}}{\I_0} \,.
\end{equation}
With
\begin{equation}
\rho \equiv \frac{M}{B_0} = M \frac{f^2}{|\langle \bar q q\rangle|}=g_A^2 \frac{\I_0}{\I_{-2}}
\end{equation}
%Note the power $g_A^2$ as opposed to the power $g_A$ in the NJL model,
%Eq.~(\ref{eq:f0NJL}). The difference has to do with the absence of a
%mass term like $ {\cal L}_m$ in the GM model. 
the result we find for the GLD coefficients reads
\begin{eqnarray}
L_1 &=& \frac{N_c}{48(4\pi)^2}\Big[(1-g_A^2)^2 \I_0 \nonumber \\
&+&4g_A^2(1-g_A^2) \I_2+2g_A^4 \I_4\Big] \,, \label{eq:L1} \\ 
L_2 &=& 2L_1 \,, \\ 
L_3 &=& -\frac{N_c}{24(4\pi)^2}
\Big[3(1-g_A^2)^2\I_0 +8g_A^4\I_4 \nonumber \\
&+&4g_A^2(3-4g_A^2) \I_2 \Big] \,, \\ 
L_4 &=& 0 \,, \\ 
L_5
&=&\frac{N_c}{2(4\pi)^2}\rho g_A^2\left[\I_0- \I_2\right] \,, \\ 
L_6 &=&
0 \,, \\ 
L_7 &=& -\frac{N_c}{24(4\pi)^2N_f}\ga \left[6\rho\I_0-\ga \I_2\right] \,, \\ 
L_8 &=&
-\frac{N_c}{24(4\pi)^2}\left[6\rho(\rho-g_A)\I_0+g_A^2 \I_2\right] \,,\\ 
L_9 &=&
\frac{N_c}{6(4\pi)^2}\left[(1-g_A^2)\I_0+2g_A^2 \I_2\right]\,, \\
L_{10} &=& -\frac{N_c}{6(4\pi)^2}\left[(1-g_A^2)\I_0+g_A^2
\I_2\right]\,, \\ 
L_{11} &=& \frac{N_c}{12(4\pi)^2}g_A^2\I_2 \,, \\ 
L_{12} &=& -\frac{N_c}{6(4\pi)^2}g_A^2\I_2 \,, \\ 
L_{13} &=& \frac{N_c}{12(4\pi)^2}\rho\I_0 = \frac{\rho}{48 M^2}
\frac{f^2}{g_A^2}\,, \\
H_0 &=& -\frac{N_c N_f}{6(4\pi)^2}M^2\I_{-2} =
-\frac{N_f}{24}\frac{f^2}{\rho} \,,
 \label{eq:H0} \\ 
H_1 &=&
\frac{N_c}{12(4\pi)^2}\left[-(1+g_A^2)\I_0+g_A^2 \I_2\right] \,,\\ 
H_2 &=& \frac{N_c}{12(4\pi)^2}\left[6\rho^2 \I_{-2}-6\rho(\rho+g_A)\I_0+g_A^2\I_2\right]
\label{eq:H2} \, , \\ 
H_3 &=& -\frac{N_c N_f}{144(4\pi)^2}\I_0 = -\frac{N_f}{576
M^2}\frac{f^2}{g_A^2} \,, \\ H_4 &=& \frac{N_cN_f}{90(4\pi)^2}\I_0 =
\frac{N_f}{360 M^2}\frac{f^2}{g_A^2} \,, \\ H_5 &=& \frac{7N_c
N_f}{720(4\pi)^2}\I_0 = \frac{7N_f}{2880 M^2}\frac{f^2}{g_A^2}
\,. \label{eq:H5}
\end{eqnarray}
We take $M=300 \,\textrm{MeV}$ and $g_A = 0.75$. For given values of
$M$ and $g_A$, the cutoff $\Lambda$ is adjusted to reproduce $f_\pi = 93.2 \,
\textrm{MeV}$. This yields
\begin{eqnarray}
&&\Lambda =  1470\,\textrm{MeV} \,,
 \quad B_0=4913\,\textrm{MeV} \,,
\nonumber \\ 
&&
{\cal I}_{-2} = 20.8 \,, \quad
{\cal I}_0 = 2.26  \,, \quad 
{\cal I}_2 = 0.922  \,, \quad 
\nonumber \\ 
&&
{\cal I}_4 = 0.995    \,.
\end{eqnarray}
The constituent chiral quark model (QC) corresponds to take $g_A = 1$
in the previous coefficients. Using the same value for $M$, we have
for this model
\begin{eqnarray}
&&
\Lambda =  828\,\textrm{MeV} \,, 
 \quad B_0=1299\,\textrm{MeV} \,,
\nonumber \\
&&
{\cal I}_{-2} = 5.50 \,, \quad
{\cal I}_0 = 1.27 \,, \quad
{\cal I}_2 = 0.781 \,, \quad
\nonumber \\
&&
{\cal I}_4 = 0.963  \,.  
\label{eq:QC}
\end{eqnarray}
The numerical values for the GLD coefficients are displayed in
Table~\ref{tab:table2}.
%\begin{eqnarray}
%L_1 &=& \frac{N_c}{24(4\pi)^2} \I_4 \,, \\
%L_2 &=& 2L_1 \,, \\
%L_3 &=& \frac{N_c}{6(4\pi)^2}
%\left[\I_2-2 \I_4\right] \,, \\
%L_4 &=& 0 \,, \\
%L_5 &=&\frac{N_c}{2(4\pi)^2}\rho\left[\I_0-\I_2\right] \,, \\
%L_6 &=& 0 \,, \\
%L_7 &=& -\frac{N_c}{24(4\pi)^2N_f}\left[6\rho\I_0-\I_2\right] \,, \\
%L_8 &=& -\frac{N_c}{24(4\pi)^2}\left[6\rho(\rho-1)\I_0+\I_2\right] \,,\\
%L_9 &=& \frac{N_c}{3(4\pi)^2} \I_2\,, \\
%L_{10} &=& -\frac{N_c}{6(4\pi)^2} \I_2\,, \\
%L_{11} &=& \frac{N_c}{12(4\pi)^2} \I_2\,, \\
%L_{12}&=&-2L_{11} \,, \\
%L_{13}&=&\frac{N_c}{12(4\pi)^2}\rho\I_0=\frac{\rho f^2}{48M^2}=\frac{1}{6}L_5 + \frac{1}{4}\rho L_9 \,, \\
%H_0 &=& -\frac{N_cN_f}{6(4\pi)^2}\J_{-1}, \\
%H_1 &=& \frac{N_c}{12(4\pi)^2}\left[-2\I_0+ \I_2\right] \,,\\
%H_2 &=& \frac{N_c}{12(4\pi)^2}\left[-6\rho^2\I_0+ \I_2\right] \,, \\
%H_3&=&-\frac{N_cN_f}{144(4\pi)^2}\I_0=-\frac{N_f f^2}{576M^2} \,, \\
%H_4&=& -\frac{8}{5}H_3 \,, \\
%H_5&=&-\frac{7}{5}H_3 \,.
%\end{eqnarray}

\subsection{NJL model} 
\label{sec:njl_res}
The GLD coefficients for this model will have two different
contributions: one coming from the quark loop and subsequent
integration of spin $1$ fields, and another coming from the
integration of spin $0$ fields. For the quark loop contribution we
have the same expressions as Eq.~(\ref{eq:L1})-(\ref{eq:H5}). The pion
weak decay constant is
\begin{equation}
f^2 = \frac{N_c}{4\pi^2}g_A M^2 \I_0 \,. \label{eq:f0NJL}
\end{equation}
Note the power $g_A$ as opposed to the power $g_A^2$ in the GM model,
Eq. (\ref{eq:f0GM}). The difference has to do with the absence of a mass term like ${\cal L}_m$ in the GM model. We use the notation
\begin{equation}
B_0 = \frac{M}{2 G_S f^2} = \frac{M}{g_A}\frac{\I_{-2}}{\I_0}\,, 
\qquad g_A = 1-2 G_V f^2 \,.
\end{equation}
With
\begin{equation}
 \rho \equiv \frac{M}{B_0}=g_A \frac{\I_0}{\I_{-2}} \,,  
\label{eq:const}
\end{equation}
the spin 0 contribution becomes
\begin{eqnarray}
L^S_3 &=&
\frac{N_c}{4(4\pi)^2}\frac{g_A^4}{\I_0}\left[\I_0-\I_2\right]^2 \,, \\
L^S_5 &=& \frac{N_c}{4(4\pi)^2}g_A^2(g_A-2\rho) \left[\I_0- \I_2\right]
\,, \\ 
L^S_8 &=& \frac{N_c}{16(4\pi)^2}(g_A-2\rho)^2 \I_0 \,, \\
L^S_{11} &=& \frac{N_c}{12(4\pi)^2}g_A^2\left[\I_0- \I_2\right] \,, \\
L^S_{13} &=& \frac{N_c}{24(4\pi)^2}(g_A-2\rho)\I_0 \,, \\ H^S_2 &=& 2
L^S_8 \,, \\ H^S_3 &=& \frac{N_c N_f}{144(4\pi)^2}\I_0 =
\frac{N_f}{576M^2}\frac{f^2}{g_A} \,.
\end{eqnarray}
The remaining $L^S_i$, $H^S_i$ vanish identically. The sum of both
contributions will give the GLD coefficients for the NJL model, which
read
\begin{eqnarray}
L_1 &=& \frac{N_c}{48(4\pi)^2}\Big[(1-g_A^2)^2 \I_0 \nonumber \\
&+&4g_A^2(1-g_A^2) \I_2+2g_A^4 \I_4\Big] \,, \label{eq:L1NJL} \\ 
L_2 &=& 2L_1 \,, \\ 
L_3 &=& -\frac{N_c}{24(4\pi)^2}
\Big[3(1-2g_A^2-g_A^4)\I_0 +8g_A^4\I_4 \nonumber \\
&+&2g_A^2\left(2(3-g_A^2)-3g_A^2 \frac{\I_2}{\I_0}\right) \I_2 \Big] \,, \\ 
L_4 &=& 0 \,, \\ 
L_5
&=&\frac{N_c}{4(4\pi)^2}g_A^3\left[\I_0- \I_2\right] \,, \\ 
L_6 &=&
0 \,, \\ 
L_7 &=& -\frac{N_c}{24(4\pi)^2N_f}\ga \left[6\rho\I_0-\ga \I_2\right] \,, \\ 
L_8 &=&
\frac{N_c}{48(4\pi)^2}g_A^2\left[3\I_0-2\I_2\right] \,,\\ 
L_9 &=&
\frac{N_c}{6(4\pi)^2}\left[(1-g_A^2)\I_0+2g_A^2 \I_2\right]\,, \\
L_{10} &=& -\frac{N_c}{6(4\pi)^2}\left[(1-g_A^2)\I_0+g_A^2
\I_2\right]\,, \\ 
L_{11} &=& \frac{N_c}{12(4\pi)^2}g_A^2\I_0 =
\frac{g_A f^2}{48M^2} \,, \\ 
L_{12} &=& -\frac{N_c}{6(4\pi)^2}g_A^2 \I_2 \,, \\ 
L_{13} &=& \frac{N_c}{24(4\pi)^2}g_A\I_0 =\frac{f^2}{96M^2}\,, \\ 
H_0 &=& -\frac{N_c N_f}{6(4\pi)^2}M^2\I_{-2} =
-\frac{N_f}{24}\frac{f^2}{\rho}\,, \\ 
H_1 &=& \frac{N_c}{12(4\pi)^2}\left[-(1+g_A^2)\I_0+g_A^2 \I_2\right] \,,\\
H_2 &=&
\frac{N_c}{24(4\pi)^2}\big[12\rho^2\I_{-2}+3g_A(g_A-8\rho)\I_0 \nonumber \\
   &&\qquad\qquad + 2g_A^2 \I_2\big] \,,  \\ 
H_3 &=& 0 \,, \\ H_4 &=& \frac{N_c N_f}{90(4\pi)^2}\I_0
=\frac{N_f}{360M^2}\frac{f^2}{g_A}\,, \\ H_5 &=& \frac{7N_c
N_f}{720(4\pi)^2}\I_0 =\frac{7N_f}{2880 M^2}\frac{f^2}{g_A}\,.
\label{eq:H5NJL}
\end{eqnarray}
The coefficients $L_1$, $L_2$, $L_4$, $L_6$, $L_7$, $L_9$, $L_{10}$, $L_{12}$, $H_0$, $H_1$, $H_4$ and $H_5$ in the GM model coincide with those of the NJL model. However, we prefer to display them explicitly for easier reference and because the expressions for $f^2$ (Eqs.~(\ref{eq:f0GM}) and (\ref{eq:f0NJL})) do not coincide. 

Note that this model reproduces the relation $L_3 = -6L_1$, provided
terms $O(N_c g_A^4)$ are neglected. We note several differences with
previous works. The values $L_1$, $L_2$, $L_3$, $L_4$, $L_5$, $L_6$,
$L_9$, $L_{10}$, $H_1$ and $H_2$ coincide with
Ref.~\cite{Bijnens:1992uz}. $L_8$ differs in two powers of $g_A$ in
the term proportional to $\I_2$. (We reproduce their results for every
separate pieces: quark loop contribution and spin 0 contribution.)

The value of $L_7$ is non-cero, if one imposes the condition
$\textrm{Det}(U)=1$ within $\textrm{SU}(N_f)$ flavor symmetry. Both
in Ref.~\cite{RuizArriola:1991gc} and \cite{Bijnens:1992uz} this term is
not obtained, in spite of the fact that in those papers it is
explicitly mentioned that one works in $\textrm{SU}(N_f)$ flavor
symmetry. The situation can be mended if one considers
$\textrm{U}(N_f)$ flavor symmetry instead where $L_7=0$. 

The coefficients $L_{1-10}$ where given in Ref.~\cite{RuizArriola:1991gc}.
The present values of $L_4$, $L_5$, $L_6$, $L_8$, $L_9$ and $L_{10}$
coincide with those \cite{RuizArriola:1991gc} where an extra erroneous
term in $L_1$ appears. $L_3$ differs from Ref.~\cite{RuizArriola:1991gc}
in all factors except for the one in $\I_4$. We correct here these
errors (see also Ref.~\cite{Bijnens:1992uz}). $H_1$ and $H_2$ did not
appear in that reference.

The coefficients $L_{11}$, $L_{12}$ and $L_{13}$, as well as
$H_{0,3-5}$, are new and are the main result of this work. They have
been also evaluated some time ago~\cite{Andrianov:1998fr} in a quark model
without scalars and vectors and more recently by
us~\cite{Megias:2004uj} in the spectral quark model.

The numerical values for these coefficients
Eq. (\ref{eq:L1NJL})-(\ref{eq:H5NJL}) are displayed in
Table~\ref{tab:table2} in two different cases: one for the generalized
SU(3) NJL model, and another one in which integration of spin 1 fields
is not considered, i.e. $g_A=1$. For the first case we take $g_A=0.606$
as a reasonable value. Using $M=300\,\textrm{MeV}$, this yields
\begin{eqnarray}
&&\Lambda = 1344 \, \textrm{MeV} \,,
 \quad B_0=4015\,\textrm{MeV} \,,
 \nonumber \\ 
&&
\I_{-2} = 17.0 \,, \quad
\I_0 = 2.10  \,, \quad
\I_2 = 0.907  \,, \quad
 \nonumber \\ 
&&
\I_4 = 0.993  \,.
\end{eqnarray}
On the other hand, for $g_A=1$, the numerical values of $\Lambda$,
$B_0$, $\rho$ and $\I_{2n}$ are identical to those for the QC model,
Eq.(\ref{eq:QC}).  The LEC's in the NJL model with $g_A=1$ and in the QC
model differ due to the scalar contributions $L^S_{3,5,8,11,13}$ and
$H^S_{2,3}$ which are not present in the QC case.

\begin{table*}
\caption{\label{tab:table2} The dimensionless low energy constants and $H_0$
compared with some reference values and other
models. The values quoted for $L_{1-10},H_{1-5}$ are to be multiplied 
by $10^{-3}$. The value quoted for $H_0$ is to be multiplied by
$10^3\,\text{MeV}^2$. }
\begin{ruledtabular}
%\begin{tabular}{cccccccccc}
\begin{tabular} {lrrrrrrrrr}
  & ChPT\footnotemark[1]  & NJL & NJL & QC & GM & SQM\footnotemark[2] & Large $N_c$\footnotemark[3] & Dual\footnotemark[2]\\  & & & ($g_A=1$) & & & (MDM)  & &  Large $N_c$\\
\hline
$L_1$  & $0.53\pm 0.25$ & $0.77$  & $0.76$  & $0.76$  & $0.78$ & $0.79$  & $0.9$ &  $0.79$ \\
$L_2$ & $0.71\pm 0.27$  & 1.54  & 1.52 & 1.52 & 1.56 &  1.58 & 1.8 &  1.58\\
$L_3$ & $-2.72\pm 1.12$  & $-4.02$  & $-2.73$ & $-3.62$  & $-4.25$ &  $-3.17$ & $-4.3$ & $-3.17$ \\
$L_4$ & 0  & 0 & 0 & 0  & 0 & 0 & 0 & 0 & \\
$L_5$ &  $0.91\pm0.15$ & 1.26  & 2.32 & 1.08 & 0.44 & $2.0\pm0.1$ & 2.1 &  3.17\\
$L_6$ &  0 & 0 & 0 & 0  & 0 & 0  & 0 & 0 & \\
$L_7$ &  $-0.32\pm0.15$ & $-0.06$  & $-0.26$  & $-0.26$  & $-0.03$  &  $-0.07\pm0.01$ & $-0.3$ & \\
$L_8$ &  $0.62\pm0.20$  & 0.65  & 0.89 & 0.46  & 0.04 & $0.08\pm0.04$  & 0.8 & 1.18 \\
$L_9$ &  $5.93\pm0.43$  & 6.31 & 4.95 & 4.95 &  6.41 & 6.33 & 7.1 &  6.33 \\
$L_{10}$&  $-4.40\pm0.70$\footnotemark[4]  & $-5.25$  & $-2.47$ & $-2.47$  &  $-4.77$ & $-3.17$ & $-5.4$ &  $-4.75$\\
$L_{11}$&  $1.85\pm0.90$\footnotemark[5]  & 1.22  & 2.01 & 1.24 & 0.82 & 1.58  & 1.6\footnotemark[5] & \\
$L_{12}$&  $-2.7$\footnotemark[5] &  $-1.06$   & $-2.47$ & $-2.47$ & $-1.64$  & $-3.17$ &  $-2.7$\footnotemark[5] &\\
$L_{13}$&  $1.7\pm0.80$\footnotemark[5]  & 1.01 & 1.01  & 0.47 & 0.22 & $0.33\pm 0.01$ & 1.1\footnotemark[5] & \\
$H_0 $   &   & $-14.6$ & $-4.67$  & $-4.67$ &  $-17.7$ &  $1.09$  &  & \\
$H_1$   &  &  $-4.01$  &  $-2.78$   & $-2.78$  &   $-4.76$  &   &   &  & \\
$H_2$   &    &  1.46   &  1.45  & 0.59 & 0.49 & $-1.0\pm0.2$  & & & \\
$H_3$   &   &  0      & 0    & $-0.50$ &  $-0.89$ &   & &  &  \\
$H_4$   &   &  1.33   & 0.80     & 0.80 &  1.43  &   & &  & \\
$H_5$   &  &  1.16   & 0.70    &  0.70 & 1.25 &    & &  & \\
\end{tabular}
\end{ruledtabular}
\footnotetext[1]{The two-loop calculation of Ref.~\cite{Amoros:2001cp}}
\footnotetext[2]{Ref.~\cite{Megias:2004uj}}
\footnotetext[3]{Ref.~\cite{Ecker:1988te}}
\footnotetext[4]{Ref.~\cite{Bijnens:2002hp,Bijnens:1996wm}}
\footnotetext[5]{Ref.~\cite{Donoghue:1991qv}}
\end{table*}

\section{Summary and conclusions} 
\label{sec:diss} 

In the present work we have calculated the low energy constants of the
energy momentum tensor. Some of these LEC's are directly determined by
the standard Gasser-Leutwyler coefficients, whereas other, $L_{11-13}$
and $H_{0,3-5}$, are new as are driven by operators not present in the
(flat space) chiral Lagrangian. Technically, the best way to proceed
is to consider QCD in a curved space-time because these allows us to
work with a single scalar, the low energy Lagrangian, rather than its
variation, the energy momentum tensor, making easier both the
computation and the imposition of symmetry restrictions. In addition
the coupling to gravity is also obtained as a byproduct. The curved
space chiral Lagrangian, contains two kinds of contributions. On the
one hand, those obtained by ``minimal coupling'' from the flat space
case, ${\cal L}^{(g)}$, on the other, non-minimal pieces containing
the Riemann curvature tensor ${\cal L}^{(R)}$. In the spirit of not
introducing new fields other than the metric tensor, we have
considered only Einstein's gravity. Further new pieces would appear by
using connections with torsion or violating metricity. As happens with
gauge couplings (e.g. the magnetic moments) the non-minimal
gravitational pieces are not fixed (nor demanded) by general
covariance of the chiral Lagrangian and to obtain them requires to
couple gravity directly to QCD quarks and gluons previously to their
integration out to yield the low energy Lagrangian.

The standard LEC's, which are expected to follow from QCD, have also
been computed in QCD-like models containing chiral quarks, with some
degree of success. Here we apply the same approach for the non-minimal
pieces. We do so for various models, namely, the constituent quark
model, the NJL model with and without vector mesons, and the
Georgi-Manohar model. This extends our previous calculation carried
out within the spectral quark model \cite{Megias:2004uj} where also a
dual large $N_c$ model was introduced. The results quoted in
Table~\ref{tab:table2} include standard and non minimal coefficients
in these models as well as ChPT calculations and the single resonance
Large $N_c$ model. The results for the LEC's look roughly quite
similar. As a rule all models and fits give the same sign for all
coefficients, with the exception of $H_0$ and $H_2$ in the SQM. For
the standard Gasser-Leutwyler coefficients $L_{1-10}$ the best overall
agreement with the two-loop ChPT calculation of \cite{Amoros:2001cp}
is achieved by the NJL model with vector mesons, for which the value
of the reduced chi-square $\chi^2/N$, $(N=10)$, is $2.2$, although
the QC and GM models give results of near quality: $2.5$ and $3.6$
respectively. For the new coefficients there are no widely accepted
values. The closest agreement with the resonance saturation and large
$N_c$ estimates of \cite{Donoghue:1991qv} for $L_{11-13}$ is provided
by the NJL model without vectors mesons, for which $\chi^2/3=0.29$, but
this is, of course, not conclusive.  We should also mention the
remarkable agreement between the prediction of the SQM for these three
coefficients and that of the chiral bosonization model of
\cite{Andrianov:1998fr}.

\appendix*

\section{A spectral quark model with $g_A \neq 1 $ ?}

One of the common features of both the NJL model with $G_V \neq 0 $
and the GM model analyzed in this paper is the presence of an axial
coupling constant for the quarks $ g_A \neq 1 $. In this appendix we
analyze the possibility of extending the Spectral Quark Model to have
a coupling similar to the GM model. Let us remind that the
SQM~\cite{RuizArriola:2001rr,RuizArriola:2003bs,RuizArriola:2003wi} is
characterized by a generalized Lehmann representation for the quark
propagator in terms of a spectral function $ \rho(\omega) $ defined on
a suitable contour on the complex mass plane, which according to the
rules devised in those works should satisfy the spectral conditions
for the positive moments
\begin{eqnarray}
\rho_n=\int d\omega \rho(\omega) \omega^n = \delta_{n0} \,, \quad n
\ge 0 \,.
\label{eq:sc} 
\end{eqnarray}
The vanishing of even moments precludes a positive definite spectral
function. These conditions arise because one demands that high energy
processes are free from logarithmic corrections as it corresponds to
an effective model at low energies. This is the case for the
electromagnetic form factor of the pion or the partonic distributions
in the pion. Gluonic radiative contributions can be implemented by
renormalization group equations using the quark model results as
initial conditions for the QCD evolution (see
e.g. Ref.~\cite{RuizArriola:2002wr}).

The physical hadronic observables turn out to depend both on the
negative moments and the so-called log-moments,
\begin{eqnarray} 
\rho_n' = \int d \omega \omega ^n \log (\omega^2/\mu^2) \,. \label{eq:logm}
\end{eqnarray} 
The set of spectral conditions, Eq.~(\ref{eq:sc}), remove the
dependence on the scale $\mu$ in Eq.~(\ref{eq:logm}) and thus prevents
the occurrence of dimensional transmutation for {\it all moments
except} $ \rho_0' $. Assuming that the quark model is defined on a
very definite renormalization scale, any low energy hadronic
observable should not depend on any additional arbitrary scale, so we
expect that $\rho_0'$ does not determine hadronic properties, and in
particular low energy constants corresponding to process involving
pseudoscalar mesons, i.e.  $L_1-L_{13} $. Actually, in our low energy
analysis of the SQM~\cite{Megias:2004uj} the zeroth log-moment,
$\rho_0' $, only contributed to non hadronic LEC's corresponding to
vacuum polarization diagrams, like the photon and graviton wave
functions and the quadratic current quark mass corrections to the
vacuum energy density. As such, this scale dependence triggered by
dimensional transmutation in the SQM is innocuous since QED and
quantum gravity also renormalize the photon and graviton wave
functions respectively.

If we assume a Spectral Quark Model a la Georgi-Manohar and consider a
low energy expansion along the lines explained in the paper we would
generate the scale dependent term
\begin{eqnarray}
{\cal L}&=&\frac{N_c}{(4\pi)^2} \frac{1}{6} \rho_0' \Big[ ( 1-g_A^2)
  \langle F_{\mu\nu}^L U F^{\mu\nu}{}^R U^\dagger \rangle \nonumber \\
  &-& i (1-g_A^2) \langle F_{\mu\nu}^L D^\mu U D^\nu
  U^\dagger + F_{\mu\nu}^R D^\mu U^\dagger D^\nu U \rangle
  \nonumber \\ 
  &-& \frac{1}{4} (1-g_A^2 )^2 \langle (D_\mu U^\dagger
  D_\nu U )^2 \rangle  \nonumber \\
  &+& \frac{1}{4}(1-g_A^2)^2 \langle (D_\mu U^\dagger D^\mu U)^2 \rangle \Big]
 \,.
\end{eqnarray} 
This terms arises from the photon wave function contribution and
because of the additivity of the $(1-g_A) $ terms and the external
spin one fields. This yields contributions to the LEC's involving
pseudoscalars which depend on the arbitrary scale generated by
dimensional transmutation. It is not clear whether this deficiency can
be attributed to the $g_A \neq 1 $ assumption or the spectral quark
model construction.

\begin{acknowledgments}

This work is supported in part by funds provided by the Spanish DGI
with grant no. BMF2002-03218, Junta de Andaluc\'{\i}a grant no. FM-225
and EURIDICE grant number HPRN-CT-2003-00311.

\end{acknowledgments}

%\bibliography{Refs}
%\bibliographystyle{h-physrev4}

\end{document}